\documentclass[aps,prl,showpacs,floatfix,twocolumn]{revtex4-1}
\usepackage{amssymb,amsfonts,amsmath}
\usepackage{graphicx,psfrag,xspace}
\usepackage{color}
\usepackage{natbib}
\usepackage{dcolumn}
\usepackage{bm}
\usepackage{subfigure}
\usepackage{setspace}
\usepackage{url}
\usepackage{xcolor}
\usepackage{version}

\newcommand{\dd}{{\rm d}}

\begin{document}


\title{Universal properties of branching random walks in confined geometries}

\author{Cl\'elia de Mulatier$^{1,2}$, Alain Mazzolo$^{1}$, and Andrea Zoia$^{1}$}
\email{andrea.zoia@cea.fr}
\affiliation{$^{1}$CEA/Saclay, DEN/DANS/DM2S/SERMA/LTSD - Gif-sur-Yvette, France\\$^{2}$Universit\'e Paris-Sud, LPTMS, CNRS (UMR 8626) - 91405 Orsay Cedex, France}

\begin{abstract}
Characterizing the occupation statistics of a radiation flow through confined geometries is key to such technological issues as nuclear reactor design and medical diagnosis. This amounts to assessing the distribution of the travelled length $\ell$ and the number of collisions $n$ performed by the underlying stochastic transport process, for which remarkably simple Cauchy-like formulas were established in the case of branching Pearson random walks with exponentially distributed jumps. In this Letter, we show that such formulas strikingly carry over to the much broader class of branching processes with arbitrary jumps, provided that scattering is isotropic and the average jump size is finite.
\end{abstract}


\maketitle  

Precisely quantifying the flow of radiation such as neutrons or photons through a structural material or a living body represents a long-standing problem in statistical physics~\cite{chandrasekhar, duderstadt, case} and is key to mastering relevant technological issues encompassing the design of nuclear reactors~\cite{bell}, light distribution in tissues for medical diagnosis~\cite{tuchin}, and radiative heat transfer~\cite{modest}, only to name a few. A fundamental question concerns the occupation statistics of the transported particles within the body when entering from the outer surface: $i)$ the distribution of the total travelled length $\ell$, which is directly proportional to the radiation flux, and $ii)$ the distribution of the number $n$ of performed collisions, which is related to the power density deposited in the traversed region~\cite{duderstadt, case}. In this respect, occupation statistics is intimately connected to the problem of the sojourn time of a random walker in a given domain~\cite{redner, weiss}, and is naturally formulated in the framework of the stochastic process underlying the evolution of the radiation field. Linear transport (where particles are fairly diluted, i.e., interact with the surrounding medium but not with each other) is modeled in terms of Pearson random walks: particles move at constant speed along straight paths of random length, interrupted by collisions with the medium, whereupon directions are randomly redistributed~\cite{duderstadt, case, grosjean}. Generally speaking, stochastic radiation transport is coupled to a birth-death mechanism (think for example of neutron multiplication in fissile materials, or photon cascades): a random number of particles may emerge from a collision, which leads to branching particle trajectories~\cite{harris, pazsit}.

In particular, branching Pearson random walks with exponentially distributed lengths stem from assuming that the traversed medium is homogeneous at the length scale seen by the walkers along their paths. If scattering centers are completely uncorrelated, the probability of occurrence of particle-medium interactions per unit length depends only on the infinitesimal travelled distance and the flights are therefore Poissonian, i.e., exponential~\cite{duderstadt, case}. In this case, the Markovian (memoryless) nature of the transport process allows resorting to the Feynman-Kac formalism~\cite{zoia1}, from which follows a set of remarkably simple Cauchy-like formulas relating the surface and volume averages of $L=\mathbb{E}[\ell]$ and $N=\mathbb{E}[n]$~\cite{Zoia2012}, namely,
\begin{align}
\langle L\rangle_{_{\scriptstyle \Sigma}} &= \eta_d \frac{V}{\Sigma} \bigg[1 + \frac{(\nu-1)}{\lambda} \langle L\rangle_{_{\scriptstyle V}}\bigg]
\label{Cauchy_formula_L}\\
\langle N\rangle_{_{\scriptstyle \Sigma}} &= \eta_d \frac{V}{\lambda \Sigma} \bigg[1 + (\nu-1) \langle N\rangle_{_{\scriptstyle V}}\bigg].
\label{Cauchy_formula_N}
\end{align}
Here, $\mathbb{E}[\cdot]$ denotes the ensemble average over realizations, $\langle \cdot \rangle_\Sigma$ the spatial average over trajectories entering the medium through the outer surface $\Sigma$ of the body, $\langle \cdot \rangle_V$ the spatial average over trajectories starting from within the volume $V$; $\lambda$ is the mean free path of the walkers, $\nu$ the average number of descendants at each collision, and $\eta_d$ a dimension-dependent constant ($\eta_2 = \pi$ and $\eta_3 = 4$). The dimensionless quantities $\langle L\rangle_\Sigma / \lambda$ and $\langle N\rangle_\Sigma $ play a prominent role, in that they allow assessing the opacity of the body, i.e., its `size' with respect to the traversing radiation flow~\cite{duderstadt, case}. Eqs.~\eqref{Cauchy_formula_L} and~\eqref{Cauchy_formula_N} generalize the elegant Cauchy formulas previously obtained for purely diffusive exponential Pearson walks~\cite{BlancoFournier2003, Mazzolo2004, Benichou2005EPL, BlancoFournier2006}. As a particular case, when $\nu=1$ the average travelled length reads
\begin{equation}
\langle L\rangle_{_{\scriptstyle \Sigma}} = \eta_d \frac{V}{\Sigma},
\end{equation}
depending only on a purely geometric ratio and not on the specific details of the process~\cite{Zoia2012}.

In many important applications of linear transport theory, including light propagation through engineered optical materials~\cite{NatureOptical, PREOptical, PREQuenched} or turbid media~\cite{davis, davis_lecture, kostinski}, neutron diffusion in pebble-bed reactors~\cite{larsen}, and radiation trapping in hot atomic vapours~\cite{NatureVapours}, the hypothesis of uncorrelated scattering centers is however deemed to fail, which thus calls for models based on non-exponential random walks. One is then naturally led to wonder whether similar general results for $L$ and $N$ can be established in such circumstances. In this Letter, we will actually show that under mild hypotheses Cauchy-like formulas~\eqref{Cauchy_formula_L} and~\eqref{Cauchy_formula_N} have a universal character, and quite surprisingly carry over to branching Pearson walks with arbitrary jumps. In doing so, we will also derive a local version of formulas~\eqref{Cauchy_formula_L} and~\eqref{Cauchy_formula_N} relating the travelled length density and the collision density at any point of the phase space.

\paragraph{General setup.} Consider an isotropic source of radiation uniformly distributed in space. Particles leaving from the source move at constant speed and undergo jumps of random length $l$ distributed according to the density $t(l)$, with finite mean free path $\lambda=\int_0^{+\infty}l \,t(l)\,\dd l$. Upon collision, with probability $q_k$ each walker gives rise to a random number $k$ of descendants, with $\nu = \sum_k k q_k$. Absorption is taken into account by the event $k=0$. Concerning the scattering distribution, we assume that the directions $\boldsymbol{\omega}$ taken by the descendants are isotropic, i.e., obey $\Omega_d^{-1}$, where $\Omega_d = 2 \pi^{d/2}/\Gamma(d/2) $ is the surface of the unit sphere in dimension $d$. Each descendant behaves independently as the progenitor particle, thus resulting in a ramified structure for the stochastic paths.

\begin{figure}[t]
\centering
\includegraphics[scale=0.4]{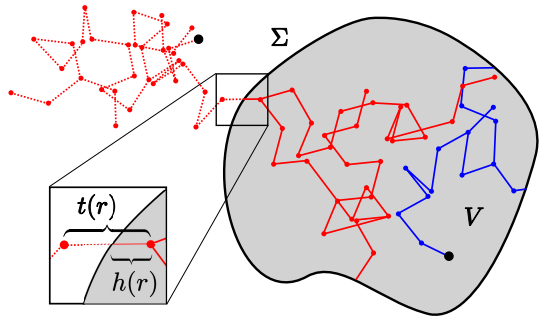}
\caption{(Color online.) Trajectories entering the body through the surface $\Sigma$ (red) and trajectories born inside the body (blue) for a branching Pearson random walk with jumps of constant size. The inset displays the first jump across $\Sigma$ for particles coming from outside.}
\label{fig1}
\end{figure}

Consider now a sub-domain of finite volume $V$ and regular surface $\Sigma$ immersed in the radiation flow. Trajectories are observed from the entrance of a single particle through $\Sigma$ until the disapperarance of the particle and all its descendants by either absorption in $V$ or escape from $\Sigma$ (see Fig.~\ref{fig1}). The previous assumptions ensure an equilibrium condition for the source, and we can safely assume that the walkers will enter the body from $\Sigma$ with a uniform distribution of entry points $\mathbf{r}_0$ and an isotropic distribution of incident directions $\boldsymbol{\omega}_0$~\cite{BlancoFournier2003, BlancoFournier2006}. This allows precisely defining the surface averages appearing in Eqs.~\eqref{Cauchy_formula_L} and~\eqref{Cauchy_formula_N}:
\begin{equation}
\langle\, f(\mathbf{r}_0, \boldsymbol{\omega}_0) \,\rangle_{_{\scriptstyle \Sigma}} = \int_\Sigma \frac{\dd \Sigma(\mathbf{r}_0)}{\Sigma} \int_{\Omega_d} \frac{\dd \boldsymbol{\omega}_0 }{\alpha_d}\, \boldsymbol{\omega}_0\cdot \mathbf{n} \, f(\mathbf{r}_0, \boldsymbol{\omega}_0)\,,
\label{S_ave_def}
\end{equation}
where $\alpha_d = 2 \pi^{(d-1)/2}/(d-1)\Gamma((d-1)/2)$ is the inward isotropic flux through a unit sphere~\cite{Benichou2005EPL, Santalo1976}. The final ingredient needed to fully characterize the particle inflow through $\Sigma$ is the density $h(r)$ of the first jump length $r$ for walkers crossing the body surface. This quantity must be proportional to the probability that the jump from outside $V$ is larger than $r$, namely, $h(r) \propto \int_r^{\infty} t(l)\dd l$. By imposing normalization and using $\lambda = \int_0^\infty \dd r \int_r^\infty t(l) \dd l$, we get the first jump length density~\cite{Mazzolo2009}
\begin{equation}
h(r) = \frac{1}{\lambda}\,\int_r^{+\infty} t(l)\,\dd l\;.
\label{first_jump}
\end{equation}
For exponential flights, we have in particular $h(r)=t(r)=\exp(-r/\lambda)/\lambda$, which is the signature of the Markovian nature of this process: trajectories crossing $\Sigma$ have no memory of their past history, so that the first jump distribution does not differ from the others.

\begin{figure}[t]
\centering
\includegraphics[scale=0.3]{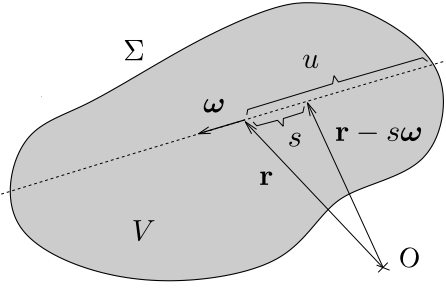}
\caption{A two-dimensional illustration of the coordinate $s$ and the distance $u=u({\mathbf r},\boldsymbol{\omega})$ from the point ${\mathbf r}$ to the surface $\Sigma$ in the direction of $-\boldsymbol{\omega}$.}
\label{fig2}
\end{figure}

\paragraph{Number of collisions.} Let us define the collision density $\psi(\mathbf{r},\boldsymbol{\omega}|\mathbf{r}_0,\boldsymbol{\omega}_0)$ such that
\begin{equation}
N(\mathbf{r}_0,\boldsymbol{\omega}_0) = \int_V \dd \mathbf{r} \int_{\Omega_d} \dd \boldsymbol{\omega}\, \psi(\mathbf{r},\boldsymbol{\omega}|\mathbf{r}_0,\boldsymbol{\omega}_0)
\label{N_def_psi}
\end{equation}
is the average number of particles having a collision within $V$, for a single walker starting from $\mathbf{r}_0$ in direction $\boldsymbol{\omega}_0$~\cite{case}. The collision density satisfies the linear Boltzmann equation~\cite{duderstadt, case}
\begin{eqnarray}
\psi({\mathbf r}, \boldsymbol{\omega}|{\mathbf r_0},{\boldsymbol{\omega}_0})&=&\nu \int_0^u \hspace{-1.6mm} \dd s \, t(s) \int_{\Omega_d} \frac{\dd \boldsymbol{\omega}' }{\Omega_d} \psi({\mathbf r}-s {\boldsymbol{\omega}} ,{\boldsymbol{\omega}'} | {\mathbf r_0},{\boldsymbol{\omega}_0})\nonumber  \\
&+& \psi_1({\mathbf r}, \boldsymbol{\omega}|{\mathbf r_0},{\boldsymbol{\omega}_0}),
\label{boltzmann_eq}
\end{eqnarray}
where $u = u({\mathbf r},{\boldsymbol{\omega}})$ is the distance from the point ${\mathbf r}$ to the surface $\Sigma$ in the direction of $-\boldsymbol{\omega}$ (see Fig.~\ref{fig2}). The quantity $\psi_1$ appearing at the right hand side of Eq.~\eqref{boltzmann_eq} is the so called first-collision density, which represents the contributions to $\psi$ due to particles having their first collision in $V$ with coordinates $\left\lbrace \mathbf{r},\boldsymbol{\omega}\right\rbrace$~\cite{case}. Let us begin by considering the trajectories coming from outside $V$ and entering the body by crossing the surface $\Sigma$ at $\mathbf{r}_0 \in \Sigma$. In this case, $\psi_1$ reads
\begin{equation}
\psi_1({\mathbf r}, \boldsymbol{\omega}|{\mathbf r_0},{\boldsymbol{\omega}_0}) = \int_0^u \dd{s}\, h(s) Q({\mathbf r}-s\boldsymbol{\omega}, \boldsymbol{\omega}),
\end{equation}
where we have set $Q({\mathbf r}, \boldsymbol{\omega}) = \delta(\mathbf{r} -\mathbf{r}_0 ) \delta(\boldsymbol{\omega} -\boldsymbol{\omega}_0 )$. Then, applying the surface average~\eqref{S_ave_def} to Eq.~\eqref{boltzmann_eq} and using the divergence theorem yields
\begin{eqnarray}
\langle\,\psi\,\rangle_{_{\scriptstyle \Sigma}}({\mathbf r},\boldsymbol{\omega}) &=&\int_0^u \hspace{-1.6mm} \dd s \, t(s) \langle\,\chi \,\rangle_{_{\scriptstyle \Sigma}}({\mathbf r}-s\boldsymbol{\omega})\nonumber\\
&-& \frac{1}{\alpha_d \Sigma} \hspace{-1mm}\int_V \hspace{-0.7mm}\dd {\mathbf r}_0 \; \hspace{-0.5mm}\nabla\hspace{-0.5mm}\Big[\boldsymbol{\omega}\,h((\mathbf{r}-\mathbf{r}_0)\cdot\boldsymbol{\omega})\Big].
\label{ave_S_boltzmann}
\end{eqnarray}
As customary in transport theory, we have here introduced the outgoing collision density $\chi ({\mathbf r}|\mathbf{r}_0,\boldsymbol{\omega}_0)= \nu\int\frac{\dd \boldsymbol{\omega}' }{\Omega_d} \psi ({\mathbf r},\boldsymbol{\omega}'|\mathbf{r}_0,\boldsymbol{\omega}_0)$, which is defined such that  $\int_V \dd \mathbf{r} \chi ({\mathbf r}|\mathbf{r}_0,\boldsymbol{\omega}_0)$ represents the average number of particles re-emitted after a collision in $V$, for a single walker starting from $\mathbf{r}_0$ in direction $\boldsymbol{\omega}_0$~\cite{case}. The integral in the second term at the right hand side of Eq.~\eqref{ave_S_boltzmann} can be explicitly computed in terms of the density $h$ and yields $-h(u)$, so that from Eq.~\eqref{ave_S_boltzmann} we are led to the following integral equation
\begin{equation}
\langle \psi \rangle_{_{\scriptstyle \Sigma}}({\mathbf r},\boldsymbol{\omega}) =\int_0^u \langle \chi \rangle_{_{\scriptstyle \Sigma}}({\mathbf r}-s\boldsymbol{\omega})t(s)\,\dd s +\frac{1}{\alpha_d \Sigma}h(u).
\label{integral_eq_psi_S}
\end{equation}
Instead of solving directly Eq.~\eqref{integral_eq_psi_S}, the idea is to relate the surface averages to the volume averages. To this aim, consider next the trajectories born within the body. In this case, $\mathbf{r}_0 \in V$ and the first-collision density reads
\begin{equation}
\psi_1({\mathbf r}, \boldsymbol{\omega}|{\mathbf r_0},{\boldsymbol{\omega}_0}) = \int_0^u \dd{s}\, t(s) Q({\mathbf r}-s\boldsymbol{\omega}, \boldsymbol{\omega}),
\end{equation}
It is convenient to introduce the volume average
\begin{equation}
\langle\, f(\mathbf{r}_0, \boldsymbol{\omega}_0) \,\rangle_{_{\scriptstyle V}} = \int_V \frac{\dd \mathbf{r}_0}{V} \int_{\Omega_d} \frac{\dd \boldsymbol{\omega}_0 }{\Omega_d} \, f(\mathbf{r}_0, \boldsymbol{\omega}_0)\,,
\label{V_ave_def}
\end{equation}
over uniform starting positions $\mathbf{r}_0 \in V$ and isotropic directions $\boldsymbol{\omega}_0$~\cite{Benichou2005EPL, Santalo1976}. Then, applying the volume average~\eqref{V_ave_def} to Eq.~\eqref{boltzmann_eq} yields the integral equation
\begin{equation}
\langle\,\psi\,\rangle_{_{\scriptstyle V}}({\mathbf r},\boldsymbol{\omega}) =\int_0^u \hspace{-1.3mm}\big[\langle\,\chi \,\rangle_{_{\scriptstyle V}}({\mathbf r}-s\boldsymbol{\omega}) + \frac{1}{V\Omega_d}\big] \, t(s) \, \dd s.
\label{integral_eq_psi_V}
\end{equation}
From inspection, it can be seen that Eqs.~\eqref{integral_eq_psi_S} and~\eqref{integral_eq_psi_V} can be both recast as a system of integral equations of the form
\begin{equation}
\left\lbrace
\begin{aligned}
& F_{_{\scriptstyle \Sigma, V}}(\mathbf{r}, \boldsymbol{\omega})=\int_0^u \big[\nu -1 + G_{_{\scriptstyle \Sigma,V}}({\mathbf r}-s\boldsymbol{\omega})\big]\, t(s)\, \dd s \nonumber \quad\\
& G_{_{\scriptstyle \Sigma, V}}(\mathbf{r})= \nu \int_{\Omega_d} \frac{\dd \boldsymbol{\omega}' }{\Omega_d} F_{_{\scriptstyle \Sigma, V}}(\mathbf{r}, \boldsymbol{\omega}'),
\label{integral_eq_psi}
\end{aligned}
\right.
\end{equation}
where $F_\Sigma(\mathbf{r}, \boldsymbol{\omega}) = \lambda \alpha_d \Sigma \langle\,\psi\,\rangle_\Sigma(\mathbf{r}, \boldsymbol{\omega}) -1$ and $F_V(\mathbf{r}, \boldsymbol{\omega})=(\nu-1)\Omega_d V \langle\,\psi\,\rangle_V(\mathbf{r}, \boldsymbol{\omega})$, respectively. Generally speaking, integral equations of this kind can not be solved explicitly, but it can be shown that their solution is unique~\cite{feller}. It thus follows the equality
\begin{equation}
\lambda \alpha_d \Sigma \langle \psi \rangle_{_{\scriptstyle \Sigma}}(\mathbf{r},\boldsymbol{\omega}) = 1+(\nu-1) V\Omega_d \langle \psi \rangle_{_{\scriptstyle V}}(\mathbf{r},\boldsymbol{\omega}).
\label{integral_equality_N}
\end{equation}
Finally, by recalling the definition in Eq.~\eqref{N_def_psi}, integrating Eq.~\eqref{integral_equality_N} over volume $V$ and over directions $\Omega_d$ and setting $\eta_d = \Omega_d / \alpha_d = \sqrt{\pi} (d-1) \Gamma((d-1)/2)/\Gamma(d/2)$ yields Eq.~\eqref{Cauchy_formula_N} as announced. Actually, the result obtained in Eq.~\eqref{integral_equality_N} is stronger than Eq.~\eqref{Cauchy_formula_N}, in that it represents a local property which is valid for any pair of coordinates $\mathbf{r}$ and $\boldsymbol{\omega}$. Observe that when $\nu =1$ Eq.~\eqref{integral_equality_N} does not depend on $\langle \psi \rangle_V$, and the corresponding surface-averaged collision density is constant over the body, namely, $\langle \psi \rangle_\Sigma(\mathbf{r},\boldsymbol{\omega}) = 1/\lambda \alpha_d \Sigma$. In this case, it follows that the average number of collisions in any sub-region $V' \subseteq V$ is simply proportional to $V'$, an ergodic-like property already exhibited for branching exponential flights~\cite{Zoia2012}.

\paragraph{Travelled lengths.} Let us define the angular flux $\varphi(\mathbf{r},\boldsymbol{\omega}|\mathbf{r}_0,\boldsymbol{\omega}_0)$ such that
\begin{equation}
L(\mathbf{r}_0,\boldsymbol{\omega}_0) = \int_V \dd \mathbf{r} \int_{\Omega_d} \dd \boldsymbol{\omega} \varphi(\mathbf{r},\boldsymbol{\omega}|\mathbf{r}_0,\boldsymbol{\omega}_0)
\label{L_def_phi}
\end{equation}
is the average length travelled in $V$, for a given walker starting from $\mathbf{r}_0$ in direction $\boldsymbol{\omega}_0$~\cite{case}. The angular flux is related to the outgoing density by~\cite{duderstadt, case}
\begin{eqnarray}
\varphi({\mathbf r}, \boldsymbol{\omega}|{\mathbf r_0},{\boldsymbol{\omega}_0})&=& \nu \hspace{-1.mm} \int_0^u \hspace{-1.mm} \dd{s}\, p_{_{\scriptstyle V}}(s) \hspace{-1.4mm} \int_{\Omega_d} \frac{\dd \boldsymbol{\omega}' }{\Omega_d} \psi({\mathbf r}-s\boldsymbol{\omega},\boldsymbol{\omega}'|{\mathbf r_0},{\boldsymbol{\omega}_0})\nonumber  \\
&+& \varphi_1({\mathbf r}, \boldsymbol{\omega}|{\mathbf r_0},{\boldsymbol{\omega}_0}),
\label{peierls_eq}
\end{eqnarray}
where $p_V(s) = 1-\int_0^s t(l)\dd l  = \lambda h(s)$ is the probability for a particle to perform a flight length larger than $s$ once emitted at a collision, and $\varphi_1({\mathbf r}, \boldsymbol{\omega}|{\mathbf r_0},{\boldsymbol{\omega}_0})$ represents the contributions to the angular flux due to uncollided particles. Consider first the trajectories coming from outside $V$ and crossing the surface $\Sigma$ at $\mathbf{r}_0 \in \Sigma$. In this case, the uncollided flux reads
\begin{equation}
\varphi_1({\mathbf r}, \boldsymbol{\omega}|{\mathbf r_0},{\boldsymbol{\omega}_0}) = \int_0^u \dd{s} \, p_{_{\scriptstyle \Sigma}}(s) Q({\mathbf r}-s\boldsymbol{\omega}, \boldsymbol{\omega}),
\label{P_delta}
\end{equation}
where $p_\Sigma(s) = 1-\int_0^s h(l)\dd l$ is the probability that the uncollided length of the walker after crossing $\Sigma$ is larger that $s$. Then, by taking the surface average of Eq.~\eqref{peierls_eq} and using the same arguments as for Eq.~\eqref{ave_S_boltzmann}, we get the integral equation
\begin{eqnarray}
& &\hspace{-11mm} \alpha_d \Sigma \langle \varphi \rangle_{_{\scriptstyle \Sigma}}({\mathbf r}, \boldsymbol{\omega}) - 1 = \nonumber\\
&=& \hspace{-2mm} \int_0^u \big[\lambda  \alpha_d \Sigma \langle \chi \rangle_{_{\scriptstyle \Sigma}}({\mathbf r}-s \boldsymbol{\omega})-1 \big]h(s)\dd s.
\label{integral_eq_phi_S}
\end{eqnarray}

Consider next the trajectories born within the body. In this case, $\mathbf{r}_0 \in V$ and the uncollided angular flux reads
\begin{equation}
\varphi_1({\mathbf r}, \boldsymbol{\omega}|{\mathbf r_0},{\boldsymbol{\omega}_0}) = \int_0^u \dd{s} \, p_{_{\scriptstyle V}}(s) Q({\mathbf r}-s\boldsymbol{\omega}, \boldsymbol{\omega}).
\end{equation}
Then, applying the volume average~\eqref{V_ave_def} to Eq.~\eqref{peierls_eq} yields the integral equation
\begin{equation}
\langle \varphi \rangle_{_{\scriptstyle V}}({\mathbf r},{\boldsymbol{\omega}})=\hspace{-1mm}\int_0^u \hspace{-1.4mm}\big[\frac{1}{V\Omega_d}+  \langle \chi \rangle_{_{\scriptstyle V}}({\mathbf r}-s \boldsymbol{\omega})\big] \lambda h(s) \dd s .
\label{integral_eq_phi_V}
\end{equation}
Equations~\eqref{integral_eq_phi_S} and~\eqref{integral_eq_phi_V} form a coupled system relating $\langle \varphi \rangle_{\Sigma,V}({\mathbf r},{\boldsymbol{\omega}})$ to $\langle \chi \rangle_{\Sigma,V}({\mathbf r})$. The integrals involving the outgoing collision densities appearing at the right hand side of Eqs.~\eqref{integral_eq_phi_S} and~\eqref{integral_eq_phi_V} can be simplified by resorting to Eqs.~\eqref{integral_eq_psi_S} and~\eqref{integral_eq_psi_V}, respectively. Then, by combining the two equations, the surface average $\langle \varphi \rangle_{_{\scriptstyle \Sigma}}({\mathbf r},{\boldsymbol{\omega}})$ can be directly solved in terms of the volume average $\langle \varphi \rangle_{_{\scriptstyle V}}({\mathbf r},{\boldsymbol{\omega}})$, from which stems the identity
\begin{equation}
\lambda \alpha_d \Sigma \langle \varphi \rangle_{_{\scriptstyle \Sigma}}(\mathbf{r},\boldsymbol{\omega}) = \lambda+(\nu-1) V\Omega_d \langle \varphi \rangle_{_{\scriptstyle V}}(\mathbf{r},\boldsymbol{\omega}).
\label{integral_equality_L}
\end{equation}
Finally, by recalling the definition in Eq.~\eqref{L_def_phi}, integrating Eq.~\eqref{integral_equality_L} over volume $V$ and over directions $\Omega_d$ yields Eq.~\eqref{Cauchy_formula_L} as announced. Similarly as for the collision densities in Eq.~\eqref{integral_equality_N}, observe that Eq.~\eqref{integral_equality_L} is valid for any pair of coordinates $\mathbf{r}$ and $\boldsymbol{\omega}$ and represents thus a result stronger than Eq.~\eqref{Cauchy_formula_L}. In particular, when $\nu =1$ Eq.~\eqref{integral_equality_L} does not depend on $\langle \varphi \rangle_V$: the corresponding surface-averaged angular flux is constant over the body, namely, $\langle \varphi \rangle_\Sigma(\mathbf{r},\boldsymbol{\omega}) = 1/\alpha_d \Sigma$, which is a purely geometrical quantity, independent of the features of the underlying random walk. In this case, the average travelled length in any sub-region $V' \subseteq V$ is simply proportional to $V'$ and thus satisfies the ergodic-like property previously established for branching exponential flights~\cite{Zoia2012}.

Formulas~\eqref{Cauchy_formula_L} and~\eqref{Cauchy_formula_N} relate surface- to volume-averaged quantities. Surface and volume terms can be also separately singled out by algebraic manipulations of Eqs.~\eqref{integral_eq_psi_S},~\eqref{integral_eq_psi_V},~\eqref{integral_eq_phi_S} and~\eqref{integral_eq_phi_V}. For the former, we have
\begin{eqnarray}
&\alpha_d \Sigma &\big[\langle \varphi\rangle_{_{\scriptstyle \Sigma}}(\mathbf{r},\boldsymbol{\omega})-\lambda\,\langle \psi \rangle_{_{\scriptstyle \Sigma}}(\mathbf{r},\boldsymbol{\omega}) \big] = \nonumber\\
&=& \hspace{-2mm}\int_0^u \big[\lambda \alpha_d \Sigma \, \langle \chi \rangle_{_{\scriptstyle \Sigma}}(\mathbf{r}-s\boldsymbol{\omega}) - 1 \big]\big[h(s)-t(s)\big] \dd s.
\label{rel_S}
\end{eqnarray}
The right hand side of Eq.~\eqref{rel_S} vanishes when $h(s)=t(s)$ for any $s$ (i.e., for exponential flights), or more generally for any class of branching Pearson walks when $\nu=1$  (for which $\langle \chi \rangle_\Sigma(\mathbf{r}) = 1/\lambda \alpha_d \Sigma$). In either case, we obtain the simple local relation $\langle \varphi \rangle_{_{\scriptstyle \Sigma}}(\mathbf{r},\boldsymbol{\omega}) = \lambda \langle \psi \rangle_{_{\scriptstyle \Sigma}}(\mathbf{r},\boldsymbol{\omega})$, from which stems also $\langle L \rangle_{_{\scriptstyle \Sigma}} = \lambda \langle N \rangle_{_{\scriptstyle \Sigma}}$. As for the latter, we get
\begin{eqnarray}
& & \Omega_d V \big[\langle \varphi \rangle_{_{\scriptstyle V}} (\mathbf{r},\boldsymbol{\omega}) - \lambda \langle  \psi\rangle_{_{\scriptstyle V}} (\mathbf{r},\boldsymbol{\omega})\big]=\nonumber \\
&=& \lambda \int_0^u \hspace{-1.3mm}\big[1+ V \Omega_d \langle \chi \rangle_{_{\scriptstyle V}}(\mathbf{r} -s \boldsymbol{\omega})\big] \big[h(s)-t(s)\big] \dd s.
\label{rel_V}
\end{eqnarray}
The quantity $1+ V \Omega_d \langle \chi \rangle_V$ is strictly positive, so that $\langle \varphi \rangle_V - \lambda \langle \psi \rangle_V $ vanishes only if $h(s)=t(s)=\exp(-s/\lambda)/\lambda$. Thus, it follows that the local relation $\langle \varphi \rangle_{_{\scriptstyle V}}(\mathbf{r},\boldsymbol{\omega}) = \lambda \langle \psi \rangle_{_{\scriptstyle V}}(\mathbf{r},\boldsymbol{\omega})$, whence also $\langle L \rangle_{_{\scriptstyle V}} = \lambda \langle N \rangle_{_{\scriptstyle V}}$, demands the Markov property of exponential flights, independent of the value of $\nu$.

\begin{figure}[t]
\centering
\includegraphics[scale=0.5]{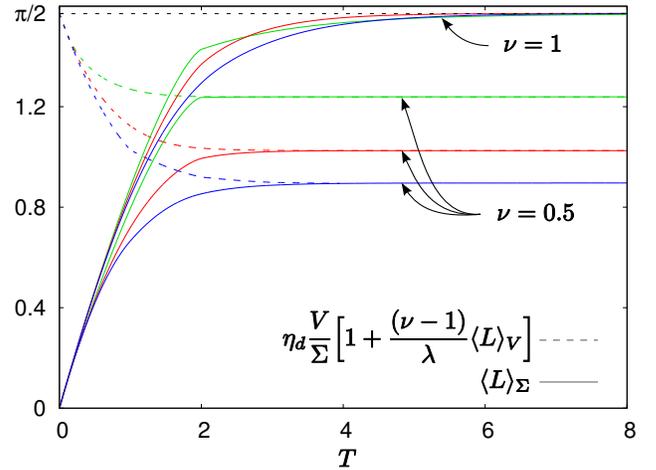}
\caption{(Color online.) The length spent in a disk of unit radius ($d=2$ and $\eta_2 V/\Sigma = \pi/2$) by branching Pearson walks with various kinds of jump distributions and various $\nu$, as obtained by Monte Carlo simulation. We consider constant flights $t(u)=\delta(u-1)$ (blue), exponential flights $t(u)=\exp(-u)$ (red), and Pareto power-law flights $t(u)=\gamma u^\gamma_m / u^{\gamma+1} $ with $\gamma=1.1$, $u_m=1/11$ and $u \ge u_m$ (green). All distributions have been normalized so that $\lambda=1$. Solid lines correspond to $\langle L \rangle_{\Sigma}$ and dashed lines to $\eta_d (V/\Sigma)[1 + (\nu-1) \langle L\rangle_{V}/\lambda]$. Both quantities are observed from the entrance of a trajectory through $\Sigma$ until the disappearance of the particle and all its descendants by either absorption in $V$ or escape through $\Sigma$. For long observation times $T$, they converge to the same value, in agreement with Eq.~\eqref{Cauchy_formula_L}. In particular, for $\nu=1$ this value is independent of the jump distribution and is given by $\eta_2 V/\Sigma=\pi/2$.}
\label{fig3}
\end{figure}

\paragraph{Discussion and perspectives.} As observed above, the form of the distribution $t(r)$ reflects the nature of the traversed medium. Neutrons and photons freely stream in the empty spaces in between obstacles, which act as scattering centers. When the mean distance between such obstacles is much larger than the average size of the obstacles, and the spatial positions are uncorrelated, then the medium (i.e., the mixture of vacuum and scattering centers) may be considered homogeneous at the scale of a mean free path $\lambda$, which ensures an exponential jump length distribution $t(r)$~\cite{duderstadt, case, PREOptical, larsen}. The hypothesis of homogeneity may break down because of spatial correlations in the scattering centers, or because of strong heterogeneities in the size of the obstacles. Optical materials with engineered obstacle sizes provide a fundamental tool for the analysis of light propagation in disordered media: when the non-scattering regions have a wide distribution spanning several orders of magnitude (fractal heterogeneity), $t(r)$ has been reported to follow a power-law decay of the kind $t(r) \sim r^{-1-\gamma}$, with $0<\gamma <2$~\cite{NatureOptical, PREOptical}. This distribution is compatible with L\'evy flights (anomalous) diffusion, whose properties are notoriously difficult to determine for confined geometries~\cite{levy}. In the case of radiative transfer in turbulent clouds, the measured jump lengths also display a power-law decay, due to the long-range correlations affecting the positions of the water droplets encountered by photons~\cite{davis, davis_lecture, kostinski}. Quenched disorder in the form of non-scattering regions similarly induces correlations between steps~\cite{PREQuenched}: this issue is central for neutron transport in pebble-bed reactors, whose core is filled with about $5\cdot 10^5$ randomly packed spheres composed of nuclear fuel and graphite, having a radius comparable to the mean free path: jump distributions appear wider than exponential, and free paths enhanced~\cite{larsen}. The description of neutron and photon propagation in such heterogeneous systems is particularly challenging, and a comprehensive theoretical framework is still missing, especially in the presence of boundaries~\cite{PREOptical, PREQuenched, larsen}. In this respect, formulas~\eqref{Cauchy_formula_L} and~\eqref{Cauchy_formula_N} contribute to the investigation of non-exponential radiation transport, in that they accommodate for arbitrary geometries and jump length distributions $t(r)$ (we require however that $\lambda < +\infty$, which for L\'evy flights would impose the restriction $\gamma >1$). For illustration, a numerical example based on Monte Carlo simulation is discussed in Fig.~\ref{fig3}.

So far, we have assumed that the surface of the body is transparent to the incoming radiation. Each re-entry from the surface (if any) is taken into account as a new trajectory, which formally corresponds to imposing absorbing boundary conditions on $\Sigma$. This is coherent with the definition given for chords traversing non-convex bodies~\cite{non_convex} and ensures the validity of the previous results for convex as well as non-convex domains. More generally, we might consider mixed boundary conditions, the surface $\Sigma$ being composed of an arbitrary combination of reflecting portions $\Sigma_r$ and absorbing portions $\Sigma_a$. Trajectories can enter the body (and escape) only through $\Sigma_a$. Collisions on $\Sigma_r$ can be indifferently modelled by assuming that the inward direction angle equals the outward direction angle (perfect reflection), or that the surface acts an isotropic diffuser~\cite{Benichou2005EPL}. In either case, by following the same strategy as above, it can be shown that any of these boundary conditions can be straightforwardly taken into account in formulas~\eqref{Cauchy_formula_L} and~\eqref{Cauchy_formula_N} by replacing the term $\Sigma$ by $\Sigma_a$~\cite{Benichou2005EPL}, which further extends the applicability of our results.

We conclude by observing that Eq.~\eqref{Cauchy_formula_N} may also prove useful in the analysis of stochastic biological populations subject to displacements, reproduction and death. In a simple random walk model of epidemics, for instance, the quantity $\langle N \rangle_\Sigma$ would provide the number of infections in a given region $V$ due to individuals coming through some frontier $\Sigma$, as long as the nonlinear effects due to the depletion of the susceptibles can be neglected~\cite{jagers, pnas}.


\begin{thebibliography}{7}

\expandafter\ifx\csname natexlab\endcsname\relax\def\natexlab#1{#1}\fi
\expandafter\ifx\csname bibnamefont\endcsname\relax
\def\bibnamefont#1{#1}\fi
\expandafter\ifx\csname bibfnamefont\endcsname\relax
\def\bibfnamefont#1{#1}\fi
\expandafter\ifx\csname citenamefont\endcsname\relax
\def\citenamefont#1{#1}\fi
\expandafter\ifx\csname url\endcsname\relax
\def\url#1{\texttt{#1}}\fi
\expandafter\ifx\csname urlprefix\endcsname\relax\def\urlprefix{URL }\fi
\providecommand{\bibinfo}[2]{#2}
\providecommand{\eprint}[2][]{\url{#2}}

\bibitem{chandrasekhar} S.~Chandrasekhar, Rev.~Mod.~Physics {\bf 15}, 1 (1943).

\bibitem{duderstadt} J.~J.~Duderstadt and W.~R.~Martin, {\em Transport theory} (J.~Wiley and sons, NY, 1979).

\bibitem{case} K.~M.~Case and P.~F.~Zweifel, {\em Linear transport theory} (Addison-Wesley, Reading, 1967).

\bibitem{bell} G.~I.~Bell and S.~Glasstone, {\em Nuclear reactor theory} (Van Nostrand Reinhold Company, New York, 1970).

\bibitem{tuchin} V.~Tuchin, {\em Tissue Optics: Light Scattering Methods and Instruments for Medical Diagnosis} (SPIE Press, 2007).

\bibitem{modest} M.~Modest, {\em Radiative heat transfer} (Academic Press, NY, 2003).

\bibitem{redner} S.~Redner, {\em A guide to first-passage processes} (Cambridge University Press, Cambridge, 2001).

\bibitem{weiss} G.~H.~Weiss, {\em Aspects and applications of the random walk} (North Holland Press, Amsterdam, 1994).

\bibitem{grosjean} C.~C.~Grosjean, Physica {\bf 19}, 29 (1953).

\bibitem{harris} T.~E.~Harris, {\em The Theory of Branching Processes} (Springer, Berlin, 1963).

\bibitem{pazsit} I.~P\'{a}zsit and L.~P\'{a}l, {\em Neutron Fluctuations: A Treatise on the Physics of Branching Processes} (Elsevier, Oxford, 2008).

\bibitem{zoia1} A.~Zoia, E.~Dumonteil, A.~Mazzolo, and S.~Mohamed, J.~Phys.~A: Math.~Theor.~{\bf 45}, 425002 (2012).

\bibitem{Zoia2012} A.~Zoia, E.~Dumonteil and A.~Mazzolo, Europhys.~Lett.~{\bf 100}, 40002 (2012).

\bibitem{BlancoFournier2003} S.~Blanco and R.~Fournier, Europhys.~Lett.~{\bf 61}, 168 (2003).

\bibitem{Mazzolo2004} A.~Mazzolo, Europhys.~Lett.~{\bf 68}, 350 (2004).

\bibitem{Benichou2005EPL} O.~B\'enichou, M.~Coppey, M.~Moreau, P.~H.~Suet, and R.~Voituriez, Europhys.~Lett.~{\bf 70}, 42 (2005).

\bibitem{BlancoFournier2006} S.~Blanco and R.~Fournier, Phys.~Rev.~Lett.~{\bf 97}, 230604 (2006).

\bibitem{NatureOptical} P.~Barthelemy, J.~Bertolotti, and D.~S.~Wiersma, Nature {\bf 453}, 495 (2009).

\bibitem{PREOptical} T.~Svensson, K.~Vynck, M.~Grisi, R.~Savo, M.~Burresi, and D.~S.~Wiersma, Phys.~Rev.~E {\bf 87}, 022120 (2013).

\bibitem{PREQuenched} T.~Svensson, K.~Vynck, E.~Adolfsson, A.~Farina, A.~Pifferi, and D.~S.~Wiersma, Phys.~Rev.~E {\bf 89}, 022141 (2014).

\bibitem{davis} A.~B.~Davis and A.~Marshak, J.~Quant.~Spectrosc.~Radiat.~Transfer {\bf 84}, 3 (2004).

\bibitem{davis_lecture} A.~B.~Davis, {\em Computational Methods in Transport, Lecture Notes in Comput. Sci. Eng.} {\bf 48}, 85 (2006).

\bibitem{kostinski} A.~B.~Kostinski and R.~A.~Shaw, J. Fluid Mech.~{\bf 434}, 389 (2001).

\bibitem{larsen} E.~W.~Larsen and R.~Vasques, J.~Quant.~Spectrosc.~Radiat.~Transfer {\bf 112}, 619 (2011).

\bibitem{NatureVapours} N.~Mercadier, W.~Guerin, M.~Chevrollier, and R.~Kaiser, Nature Physics {\bf 5}, 602 (2008).

\bibitem{Santalo1976} L.~A.~Santal\'o, {\em Integral Geometry and Geometric Probability} (Addison-Wesley, Reading, MA, 1976).

\bibitem{Mazzolo2009} A.~Mazzolo, J.~Phys.~A:~Math.~Theor.~{\bf 42}, 105002 (2009).

\bibitem{feller} W.~Feller, {\em An introduction to probability theory and its applications}, 3rd edition (Wiley, New York, 1970).

\bibitem{levy} A.~Zoia, A.~Rosso, and M.~Kardar, Phys.~Rev.~E {\bf 76}, 021116 (2007).

\bibitem{non_convex} A.~Mazzolo, B.~Roesslinger, and W.~Gille, J.~Math.~Phys.~{\bf 44}, 6195 (2003).

\bibitem{jagers} P.~Jagers, {\em Branching Processes with Biological Applications} (Wiley Series in Probability and Mathematical Statistics, London, 1975).

\bibitem{pnas} E.~Dumonteil, S.~N.~Majumdar, A.~Rosso, and A.~Zoia, Proc.~Natl.~Acad.~Sci.~USA {\bf 110}, 4239 (2013).

\end{thebibliography}
\end{document}